\documentclass{emulateapj}

\def\kms{\ifmmode{\rm km\thinspace s^{-1}}\else km\thinspace s$^{-1}$\fi}
\def\Teff{\ifmmode{T_{\rm eff}}\else $T_{\rm eff}$\fi}

\slugcomment{To appear in Astrophysical Journal.}

\shorttitle{Effect of activity on low-mass stars}
\shortauthors{Morales et al.}

\begin{document}

\title{The effect of magnetic activity on low-mass stars in eclipsing binaries}

\author{Juan Carlos Morales\altaffilmark{1}, Jos\'e Gallardo\altaffilmark{2},
Ignasi Ribas\altaffilmark{1,3}, Carme Jordi\altaffilmark{1,4},
Isabelle Baraffe\altaffilmark{5,6} and Gilles Chabrier\altaffilmark{5}}


\altaffiltext{1}{Institut d'Estudis Espacials de Catalunya (IEEC), Edif.
Nexus, C/ Gran Capit\`a, 2-4, 08034 Barcelona, Spain; morales@ieec.uab.es}
\altaffiltext{2}{Departamento de Astronom\'ia, Universidad de Chile, 
Casilla 36-D, Santiago, Chile}
\altaffiltext{3}{Institut de Ci\`encies de l'Espai (CSIC-IEEC), Campus UAB, 
Facultat de Ci\`encies, Torre C5 - parell - 2a planta, 08193 Bellaterra, Spain}
\altaffiltext{4}{Departament d'Astronomia i Meteorologia, 
Institut de Ci\`encies del Cosmos, Universitat de Barcelona, 
C/~Mart\'i i Franqu\`es 1, 08028 Barcelona, Spain}
\altaffiltext{5}{\'Ecole Normale Sup\'erieure de Lyon, CRAL (UMR CNRS 5574), 
Universit\' e de Lyon, France}
\altaffiltext{6}{School of Physics, University of Exeter, Stocker Rd, Exeter
EX4 4QL, UK}

\begin{abstract}
In recent years, analyses of eclipsing binary systems have unveiled differences
between the observed fundamental properties of low-mass stars and those
predicted by stellar structure models. Particularly, radius and effective
temperatures computed from models are $\sim 5-10$\% lower and $\sim 3-5$\%
higher than observed, respectively. These discrepancies have been attributed to
different factors, notably to the high levels of magnetic activity present on
these stars. In this paper, we test the effect of magnetic activity both on
models and on the observational analysis of eclipsing binaries using a sample
of such systems with accurate fundamental properties. Regarding stellar models,
we have found that unrealistically high spot coverages need to be assumed to
reproduce the observations. Tests considering metallicity effects and missing
opacities on models indicate that these are not able to explain the radius
discrepancies observed. With respect to the observations, we have tested the
effect of several spot distributions on the light curve analysis. Our
results show that spots cause systematic deviations on the stellar radii
derived from light curve analysis when distributed mainly over the stellar
poles. Assuming the existence of polar spots, overall agreement between models
and observations is reached when $\sim 35$\% spot coverage is considered on
stellar models. Such spot coverage induces a systematic deviation in the radius
determination from the light curve analysis of $\sim 3$\% and
is also compatible with the modulations observed on the light curves of these
systems. Finally, we have found that the effect of activity or rotation on
convective transport in partially radiative stars may also contribute to
explain the differences seen in some of the systems with shorter orbital
periods. 
\end{abstract}


\section{Introduction}
\label{sec:introduction}

Analyses of double-lined eclipsing binary stars (hereafter EBs) have provided
fundamental properties of stars on the low-mass end of the Main Sequence. Such
EB systems have revealed to be the best candidates to test stellar structure
models since they can potentially yield masses and radii for M-type stars with
uncertainties below $\sim 1$\% \citep{Morales2009a}. However, results coming
from a variety of independent studies indicate that differences exist between
the observed properties and those predicted by models; for the same mass ($M$),
models predict $\sim 5-10$\% smaller radius ($R$) and $\sim 3-5$\% larger
effective temperatures (\Teff) than observed for stars below 1~M$_{\odot}$.  In
contrast, the stellar luminosity for these binary components is in agreement
with those of single stars and also with theoretical models \citep[see][for a
review]{Ribas2006}.

Several authors have argued that activity could be the cause of discrepancies
found between models and observations of these systems
\citep{Torres2002,LopezMorales2005,Torres2006,LopezMorales2007,Ribas2008}.
Further support to this hypothesis is lent by similar radius discrepancies
found when comparing single magnetically active late-type stars with their
inactive counterparts \citep{Morales2008}, and by the results of
interferometric observations of single inactive stars \citep{Demory2009}.
Until now, all the known low-mass EBs are close binaries with periods well
below 10 days, thus they are expected to be synchronized \citep{Mazeh2008},
i.e., with their rotation tidally locked to the orbital motion, and 
therefore they are fast rotators ($v_{\rm rot} \gtrsim 10$~km~s$^{-1}$). Such
fast rotation in presence of magnetic fields could trigger high levels of
activity that may modify the structure of these stars in some respects.
Observationally, the high activity of these stars is usually confirmed by
saturated X-ray emission ($\log L_{X}/L_{bol}\sim-$3), strong chromospheric
emission in the H$\alpha$ Balmer line, flares, and modulations on the light
curves caused by the presence of starspots on the stellar surface.

Over the past decades, substantial progress towards the understanding of the
structure and evolution of low-mass stars has been made, also including
activity effects. \citet{Chabrier2007} showed that the introduction of
rotation and/or magnetic field effects on the models of \citet{Baraffe1998}
could explain the discrepancies between the observed and theoretically
predicted mass-radius relationship of EBs. Specifically, they presented two
scenarios considering: {\em (1)} that the effect of magnetic field and rotation
alter the efficiency of convective energy transport, which can be modelled by
setting the mixing length parameter ($\alpha$ from Mixing Length Theory) to
lower values than those used for solar models; and {\em (2)} that the stellar
magnetic activity present on these objects can be associated with the
appearance of dark spots covering the radiative surface, which can be modelled
by assuming a new stellar luminosity, $\hat{L}\propto \left( 1 -\beta \right)
\hat{R}^{2} \hat{T}_{\rm eff}^{4}$, where $\hat{R}$ and $\hat{T}_{\rm eff}$ are
the modified stellar radius and effective temperature, respectively, and
$\beta$ is the factor of spots blocking the outgoing luminosity due to their
lower temperature (see Section~\ref{sec:lc} for further details on this
parameter).

The results of \citet{Chabrier2007} show that both these scenarios predict
larger radius than standard stellar models, but while the effect of spots is
significant over the entire low-mass domain, the effect of convection is
relatively smaller for fully convective stars ($M \le 0.4 M_\odot$) because
convection is nearly adiabatic and changing the mixing length parameter has a
modest impact on their stellar structure. Since radius discrepancies have been
reported over a wide range of masses, this may be indicating that either the
contribution of spots or a combination of spots and activity effect on
convective transport are needed to reconcile models with observations.

New accurate fundamental properties of low-mass EBs, and especially the case of
CM~Dra \citep{Morales2009a}, providing precise characterization of two fully
convective stars, makes it worthwhile to carry out a new thorough analysis of
the agreement between observation and theory using the scenario suggested in
\cite{Chabrier2007}. In Section~\ref{sec:sample} we briefly describe the
discrepancies between stellar structure models and current observations of
low-mass stars and we select the sample of best known EB systems to constrain
stellar models. In Section~\ref{sec:lc} we analyze the effect of the
photospheric spots on EB light curves and on the parameters fitted to these
curves. Critical comparison with models including activity effects is discussed
in Section~\ref{sec:models}, and finally, the conclusions are presented in
Section~\ref{sec:conclusions}.

\section{Low-mass EBs compared to models}
\label{sec:sample}

\begin{figure*}[t]
\centering
\includegraphics[width=\textwidth]{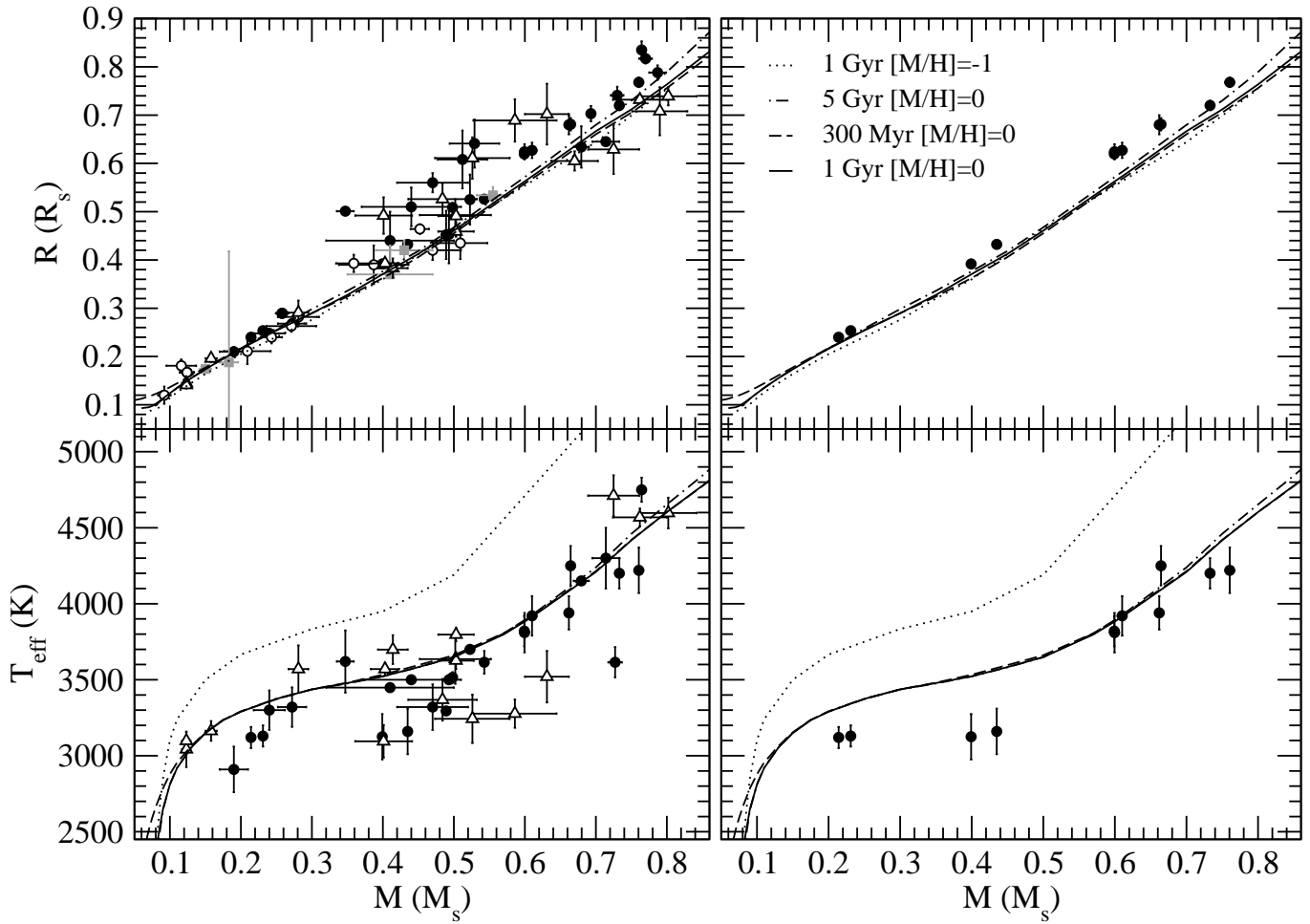}
\caption{Different sets of theoretical $M-R$ and $M-\Teff$ relationships
compared with observational values of low-mass stars. Left: Low-mass stars in
double-lined EB systems (filled circles), single-lined EB systems (open
circles), M stars in post-common envelope binaries (grey squares) and
late-type stars with radius measured interferometrically (open triangles).
Right: Same plots showing the EBs included in our analysis (see text and
Table~\ref{tab:best_EBs}). Theoretical models with different sets of
parameters as labelled are from \citet{Baraffe1998}.}
\label{fig:best_EBs}
\end{figure*}

\begin{deluxetable*}{lcccccccc}
\tablewidth{0pc}
\tabletypesize{\footnotesize}
\tablecaption{Properties of observed EBs included in our sample.\label{tab:best_EBs}}
\tablehead{\colhead{ } & \colhead{$P$} & \colhead{$M$}        & \colhead{$R$}        & \colhead{\Teff} & \colhead{ } & \colhead{Age} & \colhead{$v_{\rm rot,sync}$} & \colhead{ } \\
\colhead{EB} & \colhead{(days)} & \colhead{(M$_{\odot}$)} & \colhead{(R$_{\odot}$)} & \colhead{(K)} & \colhead{$[\rm M/\rm H]$} & \colhead{Gyr} & \colhead{(kms$^{-1}$)} & \colhead{Ref.\tablenotemark{a}}}
\startdata
V818 Tau B\dotfill                & 5.61 & 0.7605$\pm$0.0062 &  0.768$\pm$0.010  & 4220$\pm$150    &\phs0.13 & 0.6\phn   &   6.95$\pm$0.09     & 1 \\
IM Vir B\dotfill                  & 1.31 & 0.6644$\pm$0.0048 &  0.681$\pm$0.013  & 4250$\pm$130    & $-$0.28 & 2.4\phn   &  26.12$\pm$0.50\phn & 2 \\
NGC2204-S892 A\dotfill            & 0.45 &  0.733$\pm$0.005  &  0.720$\pm$0.010  & 4200$\pm$100    &   --    &  --       &   80.9$\pm$1.1\phn  & 3  \\
NGC2204-S892 B\dotfill            &      &  0.662$\pm$0.005  &  0.680$\pm$0.020  & 3940$\pm$110    &         &           &   76.4$\pm$2.2\phn  &    \\
GU Boo A\dotfill                  & 0.49 & 0.6101$\pm$0.0064 &  0.627$\pm$0.016  & 3920$\pm$130    & $\sim$0 & 1.0\phn   &   64.7$\pm$1.6\phn  & 4  \\
GU Boo B\dotfill                  &      & 0.5995$\pm$0.0064 &  0.624$\pm$0.020  & 3810$\pm$130    &         &           &   64.4$\pm$1.6\phn  &    \\
YY Gem A \& B\dotfill             & 0.81 & 0.5992$\pm$0.0047 & 0.6194$\pm$0.0057 & 3820$\pm$100    & $\sim$0 & 0.4\phn   &  38.88$\pm$0.36\phn & 4  \\
CU Cnc A\tablenotemark{b}\dotfill & 2.77 & 0.4349$\pm$0.0012 & 0.4323$\pm$0.0055 & 3160$\pm$150    & $\sim$0 & 0.32      &   8.02$\pm$0.10     & 4  \\
CU Cnc B\tablenotemark{b}\dotfill &      & 0.3992$\pm$0.0009 & 0.3916$\pm$0.0094 & 3125$\pm$150    &         &           &   7.30$\pm$0.17     &    \\
CM Dra A\dotfill                  & 1.27 & 0.2310$\pm$0.0009 & 0.2534$\pm$0.0019 & 3130$\pm$70\phn &$-1$/$-0.6$& 4.1\phn &  10.02$\pm$0.08\phn & 4  \\
CM Dra B\dotfill                  &      & 0.2141$\pm$0.0008 & 0.2398$\pm$0.0018 & 3120$\pm$70\phn &         &           &   9.51$\pm$0.07     &    \\
\enddata
\tablenotetext{a}{References: 1. \citet{Torres2002}; 2. \citet{Morales2009b}; 3. \citet{Rozyczka2009}; 4. \citet{Torres2009}}
\tablenotetext{b}{\Teff\ could be underestimated due to the presence of circumbinary dust.}
\end{deluxetable*}

The properties describing the evolutionary phase of a star are its mass,
chemical composition and age. Given these three properties, radius and
effective temperature are determined from stellar evolutionary models.
Therefore, models could be tested comparing the theoretical $M-R$ and
$M-$\Teff\ relationships with observational values. In
Figure~\ref{fig:best_EBs} such comparisons are shown for the models of
\citet{Baraffe1998} and the full sample of low-mass stars with known physical
properties. This figure includes properties derived from double-lined and
single-lined EBs, interferometric radius measurements and also from analysis of
M-type stars in post-common envelope binaries, although some of
these measurements are not independent of theoretical models or calibrations.
In post-common envelope components, the systems may not be reliable
representations of the stars described by models due to interactions suffered
during the common envelope phase.

In Figure \ref{fig:best_EBs}, a clear difference between stars below and
above $\sim 0.35$~M$_{\odot}$ is apparent. On the lower end, radii are much
closer to theoretical models and less scattered, while on the upper end, the
scatter is larger and there seems to be a larger difference with models as
well. This was already noticed in previous analyses \citep{Ribas2006} and it
may be consistent with the boundary between fully convective stars ($M \lesssim
0.35$~M$_{\odot}$) and partially radiative stars. Considering these two
regimes, differences between all known measured radii of low-mass stars and the
1~Gyr solar metallicity model isochrone are 5.7\% and 8.1\% for fully
convective and partially radiative stars, respectively. The dispersion of these
differences is 3.1\% and 12.2\%, respectively, thus illustrating the scatter
difference mentioned above. In the case of effective temperatures the
differences are $-$1.9\% and $-$1.6\%, respectively, for this sample.

To test stellar structure models, masses and radii of stars need to
be known with accuracies down to the few percent level
\citep{Andersen1991,Torres2009}. As mentioned earlier, EB systems provide the
best opportunity to measure these fundamental stellar properties with such
accuracy, as clearly visible in Figure~\ref{fig:best_EBs}. Thus, for our
analysis we have selected all known low-mass stars with accuracies better than
3\%. As reported recently by \cite{Torres2009}, special attention should be
paid both on the precision and the accuracy of the measures, and so we have
excluded from our sample systems with error estimations not clearly described.
Furthermore, we have only selected main sequence systems with ages below 5~Gyr
and masses below 0.8~M$_{\odot}$ to restrict radius differences due to
evolution below 2\% and also to minimize the effect of mixing length
differences between low-mass and solar mass models. The mechanical and thermal
properties for each component of the selected systems are listed in
Table~\ref{tab:best_EBs}, as well as the age and metallicity when known.
Comparison of this sample with models is shown on the right panels of
Figure~\ref{fig:best_EBs}. For the subsample of Table~\ref{tab:best_EBs}, the
comparison with the 1~Gyr solar metallicity model gives discrepancies with
observations of 5.8\% and $-$6.1\% in radii and effective temperatures,
respectively, for fully convective stars (CM~Dra), and 7.8\% and $-$3.0\%,
respectively, for partially radiative stars.

Besides magnetic activity, several model parameters such as metallicity
or opacities \citep{Berger2006,Casagrande2008} have been proposed to explain
the discrepancies between theory and observations. Our calculations indicate
that metallicity produces differences between $\sim 4$\% and $\sim 20$\% on the
radius and effective temperature values predicted by models, respectively, when
comparing $[\rm m/\rm H]=0.0$ and $[\rm m/\rm H]=-1.0$ models (solid and dashed
line, respectively, in Figure~\ref{fig:best_EBs}). However, none of the systems
is known to have such low metallicity, except for the case of CM~Dra, for which
its metallicity is not well established, and NGC2204-S892, for which there is
no determination. Besides, radius differences are lower than needed to
reproduce observations even with metallicities as low as $[\rm m/\rm H]=-1.0$.
Therefore, metallicity effects are expected to be not significant in our
sample. In the case of effective temperature comparisons, where the effect of
metallicity is more important, EB light curves only provide accurate values of
the temperature ratio for the components. The individual effective temperatures
are usually computed using different spectrophotometric calibrations for each
system, thus introducing the potential for systematic errors. Furthermore, some
EB systems may have peculiarities, such as CU~Cnc, which is supposed to have a
circumbinary dust disk that may affect the temperature determination. Thus, 
to investigate the agreement between models and observations we will primarily
make use of the $M-R$ relationship, which are both fundamentally determined
quantities.

Recently, the possibility of underestimated opacities has been raised in
the context of the Standard Solar Model. An increase of opacities near the
bottom of the convective zone has been shown to reconcile helioseismic data of
the Sun with models using the new chemical composition
\citep{Bahcall2004,Bahcall2005}. These results and others found in the
literature invoking problems with the atmospheric opacities as a source of the
discrepancies present in observed radii of EB compared with theoretical models
prompted us to perform different tests to analyze the effect of atmospheric
opacity on the resulting stellar radius. Using the standard Lyon group stellar
evolutionary models \citep{Baraffe1998}, an overall massive increase by a
factor of $\sim 10$ in the opacity affects the radius by just 4\% (and even this
difference is smaller below the fully convective boundary), as already shown by
\citet{Chabrier1997}. To obtain a $\sim 10$\% larger radius the opacities of
eclipsing binaries would have to be increased to unrealistically high values.
Missing elemental opacities do not seem to be enough to explain the $\sim
5-10$\% radius disagreement between models and observations. Moreover, both
metallicity and opacity effects should also be present on single stars, which
are known to be well described by models, as recently confirmed by
\cite{Demory2009} from the comparison of interferometric radius measurements
of inactive low-mass stars with model predictions. Further, the authors did
not find a correlation between metallicity and radius differences thus not
confirming the results from Berger et al. (2006). Further, similar radius
differences as those reported here are known between single magnetically
active stars and their inactive counterparts \citep{Morales2008}, thus
strengthening the activity hypothesis.

\section{Determination of physical properties of EBs}
\label{sec:lc}

The conclusions of the previous section reinforce magnetic activity as the most
reasonable explanation for the discrepancies found between stellar models and
observations of low-mass stars. Before embarking on a careful fine-tuning of
model parameters to reach agreement with observations it is worth evaluating
whether our observed eclipsing binary parameters could be affected by the
strong magnetic activity of the components. Time-variable brightness changes are
a common signature of the presence of activity on low-mas stars both on single
and eclipsing binaries. In the latter case, the modulation present on the light
curves is combined with the eclipsing variability and, therefore, starspots
must be taken into account to derive accurate stellar fundamental properties.
In this section we study the possibility that starspots could be responsible
for systematic effects inherent to the light curve and radial velocity analyses
in the case of the very active low-mass eclipsing binaries. We analyze whether
the fundamental properties determined from the classical modelling could be
biased because of the presence of starspots in different geometries.

Absolute physical properties of EBs are obtained from the combination of the
results of light and radial velocity curve analyses with modelling codes such
as the widely-used Wilson-Devinney \citep[hereafter WD;][]{WD1971}. Light
curves yield properties such as the inclination of the orbit ($i$), the
ratio of effective temperatures (\Teff$_{2}$/\Teff$_{1}$), the luminosity ratio
in the light curve bandpass ($L_{2}$/$L_{1}$) and the surface potentials
($\Omega_{1}$ and $\Omega_{2}$), which are related to the radii relative to the
semimajor axis ($r_{1}$ and $r_{2}$). Radial velocity curves provide values for
the mass ratio ($q$) and projected semi-major axis ($a\sin i$). The combination
of the two modelling procedures yields the absolute properties of the
components independently of models or distance calibrations. Radiative
parameters, such as limb darkening and gravity darkening, relevant for light
curve analyses, are usually taken from theory.

The presence of spots on the surface of the components of an EB system cause
perturbations on both the light and radial velocity curves. The most prominent
effect is on light curves, in which modulations in the out-of-eclipse phases
appear due to the transit and occultation of spots according to the orbital and
spin motions. To appropriately derive physical properties of the components,
spot effect must be taken into account when analysing these curves. The WD code
introduces spots on the modelling assuming that they are circular and that they
have an uniform temperature ratio with respect to the photosphere. With this
simple model, properties of spots, such as their location in co-latitude
($\theta_{s}$, measured from North pole) and longitude ($\phi_{s}$), the
angular size ($r_{s}$) and the temperature contrast with the photosphere, i.e.,
$\frac{\Teff_{\rm ,s}}{\Teff}$, where \Teff\ and \Teff$_{\rm ,s}$
are the effective temperatures of the photosphere and the spots, respectively,
can be fitted. The values of $\phi_{\rm s}$ and $r_{\rm s}$ are well
constrained by the central phase of the modulation and its duration and
amplitude, respectively, however the amplitude is also dependent on
$\theta_{\rm s}$ and $\frac{\Teff_{\rm ,s}}{\Teff}$, so these parameters are
strongly correlated with the size of spots. This model can reproduce the
presence of complex and irregular spot groups by equivalent circular spots with
an average temperature.

The total luminosity of a spotted star is the addition of the contribution of
the spots at an effective temperature \Teff$_{,\rm s}$ and the immaculate
surface at \Teff. If we consider $S$ and $S_{\rm s}$ to be the surface of the
star and that covered by spots, respectively, the total luminosity is given by:
\begin{equation}
\label{eq:beta_exp}
L=\left(S-S_{\rm s}\right) \sigma \Teff^{4} +
S_{\rm s} \sigma \Teff_{,\rm s}^{4}.
\end{equation}
Comparing this equation with the formalism introduced by \citet{Chabrier2007},
it can be concluded that the $\beta$ parameter is given by:
\begin{equation}
\label{eq:beta_ST}
\beta=\frac{S_{\rm s}}{S}\left[1-\left( \frac{\Teff_{\rm ,s}}{\Teff} \right)^{4} \right].
\end{equation}
This equation indicates that the $\beta$ parameter is related with the
properties of spots. For completely dark spots (\Teff$_{\rm ,s}=0$), the term
in brackets in Equation~\ref{eq:beta_ST} is exactly one, therefore, $\beta$ is
a measure of the fraction of stellar surface covered by dark spots as defined
by \cite{Chabrier2007}. However, in the realistic case that spots are not
completely dark the term in brackets is less than one and $\beta$ is lower than
the fraction of spotted surface. For a given \Teff\ and an estimation of the
contrast of spots, Equation~\ref{eq:beta_ST} provides the ratio of surface
covered by spots when $\beta$ is known, or vice versa.

Spot modelling results have been published for several of the best-known EBs
\citep{Torres2002,Ribas2003,LopezMorales2005,Morales2009a,Morales2009b}
reporting both cold an hot spots with radii between $\sim 9$~deg and
$\sim 90$~deg and with temperature factors ($\Teff_{,s}/\Teff$) down to 0.86.
Hot spots can be interpreted as photosphere regions surrounded by large cool
spots. Considering the spot parameters given in the above references for the
cases of IM~Vir, GU~Boo, YY~Gem, CU~Cnc, and CM~Dra, we find $\beta$ values 
of up to 0.1 according to Equation~\ref{eq:beta_ST}. This value is much smaller
than the range between 0.3 and 0.5 suggested by \cite{Chabrier2007}.  It must
be mentioned, though, that since the relevant measures in EB light curve
analyses are differential magnitudes, the photometric variations used to derive
the spot parameters are not sensitive to the total surface covered by spots but
to the contrast between areas with different effective temperatures. For
instance, an evenly spotted star would not show significant light curve
variations. It is thus possible that EBs could be more heavily spotted than
simple photometric variations indicate and permit $\beta$ values higher than
0.1. A check of the photometric variations that could produce such higher
values of $\beta$ and their effect on derived parameters from fits are
described in the following sections.

\subsection{Modulation on light curves}
\label{subsec:lc_modulation}

For M-type stars, an assumption of \Teff$\sim$3500~K and spots about
$\sim 500$~K cooler than the photosphere and $\beta$ values between 0.3--0.5
corresponds to a fractional surface area of 65--100\% covered by spots
(as shown by Equation~\ref{eq:beta_ST}). To test the consistency between the
theoretical modulations expected from different $\beta$ values and the spot
modulations observed on EB light curves, we simulated EB systems with
components randomly spotted and used the WD code to generate their light
curves. Such theoretical light curves were subsequently compared with those
observed in EBs.

We have developed a code to randomly place spots on the surface of stars. We
assumed a uniform longitude distribution and tried different distributions
over latitude. \cite{Granzer2000} calculate the probability of spot appearance
for different latitudes as a function of spin period and mass. Their results
show that for a star like the components of GU~Boo (see
Table~\ref{tab:best_EBs}), spots are formed in a band from 25~deg to 55~deg of
co-latitude. Both bands (one per hemisphere) represent about one third of the
stellar surface, which is not enough to simulate spot coverages with $\beta$
greater than 0.15 (assuming spots 500~K cooler than the photosphere). Doppler
tomography analyses have revealed the prominent existence of polar spots on
active stars (\citeauthor{Jeffers2007} \citeyear{Jeffers2007};
\citeauthor{Washuettl2001} \citeyear{Washuettl2001}; see also
\citeauthor{Strassmeier2009} \citeyear{Strassmeier2009}, and references
therein). Thus, we decided to extend the distribution to
all stellar latitudes. To use similar distributions as those given in
\cite{Granzer2000} we considered a linear probability from the pole to 40~deg
and from 45~deg to the equator with a plateau between 45~deg and 40~deg
(hereafter Distribution 1). This roughly mimics distributions of
0.6~M$_{\odot}$ rapidly-rotating stars. For comparison we considered also a
bilinear distribution from the pole to 70~deg with a peak at 25~deg. This is
similar to the 0.4~M$_{\odot}$ distribution in \cite{Granzer2000} for rapidly
rotating stars but extended to cover $\beta$ factors up to 0.3 (hereafter
Distribution 2). Finally, we considered a completely uniform distribution.

\begin{figure}[t]
\centering
\includegraphics[width=\columnwidth]{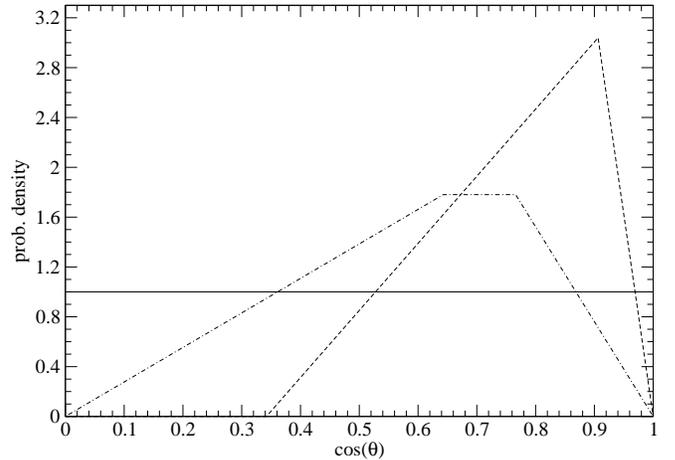}
\caption{Probability density functions over $\cos \theta$ used to simulate
spotted stars. The equator corresponds to $\theta=90$~deg. The uniform
distribution (solid line), Distribution 1 (dot-dashed line) and Distribution 2
(dashed line) are described in the text.}
\label{fig:distributions}
\end{figure}

These distributions are shown in Figure~\ref{fig:distributions} and it 
illustrates that Distribution 2 is more concentrated towards the poles.
Additionally, to have reasonable computing time using the WD code, we limited
the number of spots by imposing that the centers of two spots could not be
closer than half their radius, i.e., each spot contributes at least $\sim 30$\%
of its surface to the total spot coverage.

The simulation procedure was executed as follows. We considered spots with a
temperature contrast with the photosphere of 0.85, even on overlapping
areas (i.e., the temperature contrast in the area where two spots overlap is
also 0.85). This value constrains $\beta$ between 0 and $\sim 0.5$, the latter
meaning a completely spotted stellar surface. In the case of Distribution 2 the
upper value of $\beta$ is $\sim 0.35$. We chose 10~deg as spot radius in line
with large spot groups measured on the Sun. Then, for each latitude
distribution we simulated 25 light curves with spots randomly distributed over
the surface of the components for different values of $\beta$. From these
simulated light curves we measured the peak-to-peak variations on the
out-of-eclipse phases and computed the mean value of the 25 light curves. This
procedure was done using the physical properties of GU~Boo as template and
using $R$~band light curves. However, since the effect of absolute masses or
mass ratio on the light curves is negligible, this should be representative of
the entire low-mass star domain.

\begin{figure}[!t]
\centering
\includegraphics[width=\columnwidth]{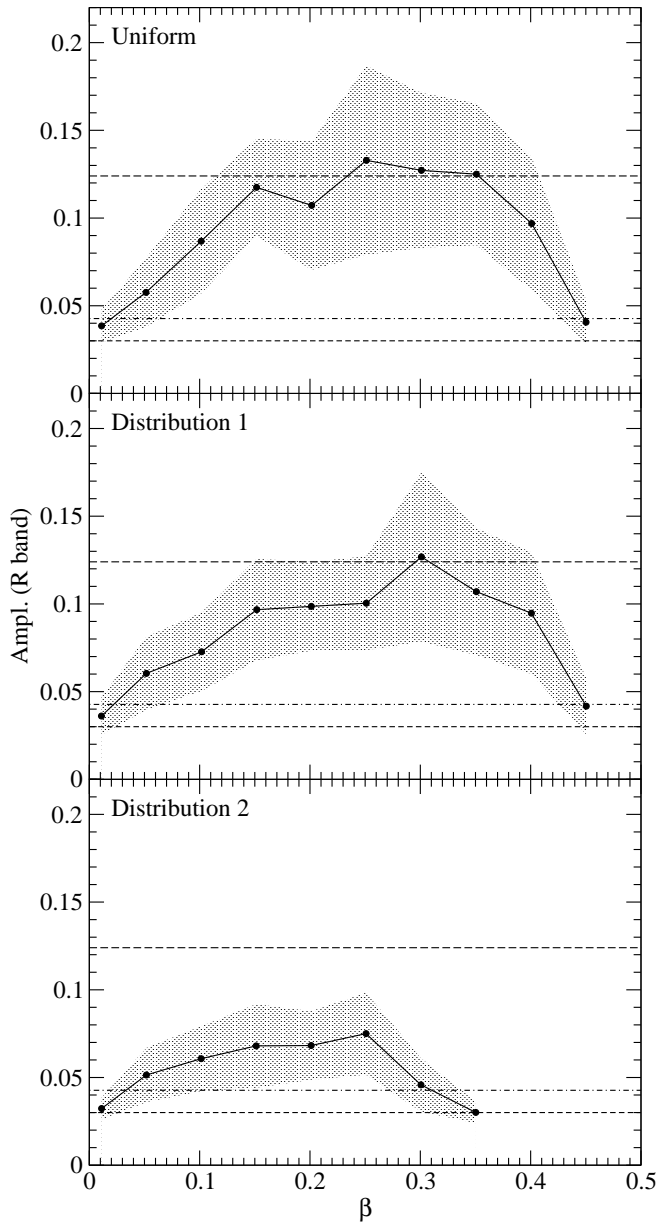}
\caption{Amplitude of the modulations produced by different $\beta$ spot
scenarios using different spot distributions on $R$ band light curves. Each
point is the mean value of 25 realizations and the shaded area shows the
standard deviation of these modulations. The values reported in the literature
for CM~Dra (dashed line), YY~Gem (dot-dashed line) and GU~Boo (long-dashed
line) are shown for comparison.}
\label{fig:modulation_comparison}
\end{figure}

\begin{figure}[!t]
\centering
\includegraphics[width=\columnwidth]{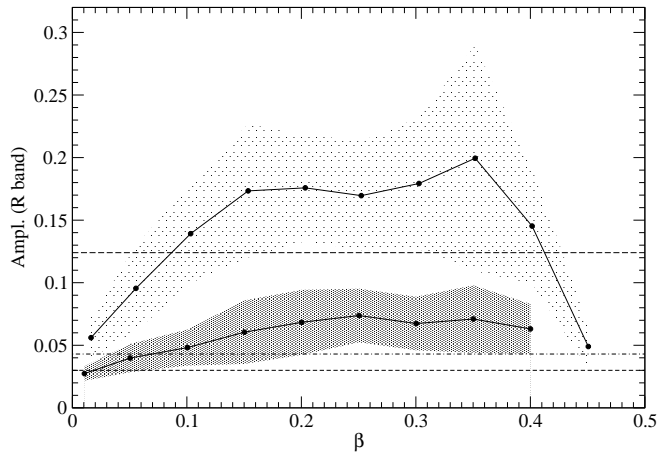}
\caption{Same as Figure~\ref{fig:modulation_comparison} for the uniform
distribution comparing the cases of simulations using spots with radius of
15~deg (light shade) and 5~deg (dark shade).}
\label{fig:modulation_r}
\end{figure}

The values of the peak-to-peak modulations found with these simulations were
later compared with the real values for the best-known EBs.
Figure~\ref{fig:modulation_comparison} shows the peak-to-peak modulations
resulting from simulations for each distribution compared with a few
observational values as reference. Large modulations are better explained by
mild spot coverages assuming a uniform distribution or Distribution 1, while
low spot signals are better reproduced by Distribution 2. However, the
amplitude of these modulations depends on the size and thermal properties of
the spots as shown on Figure~\ref{fig:modulation_r}. It is clear
that the observed modulations could be explained with any of the distributions
when considering the different spot sizes and the intrinsic dispersion of the
modulation. Therefore, low amplitudes such as those
of YY~Gem and CM~Dra could be recovered with a mild spot coverage with small
size or low temperature contrast spots. Besides, the variability of the spot
properties should also be taken into account since it could induce different
modulations on different epochs as has been reported for the case of $V$ band
light curves of YY~Gem, showing amplitudes of 0.09~mag and 0.055~mag
in different seasons \citep{Torres2002}.

Several other distributions, such as linear, quadratic and square root from
equator to poles were tested yielding results between the uniform and
Distribution 1 cases.

\subsection{Systematics on light curves due to spots}
\label{subsec:lc_systematics}

\begin{figure*}[!t]
\centering
\includegraphics[width=\columnwidth]{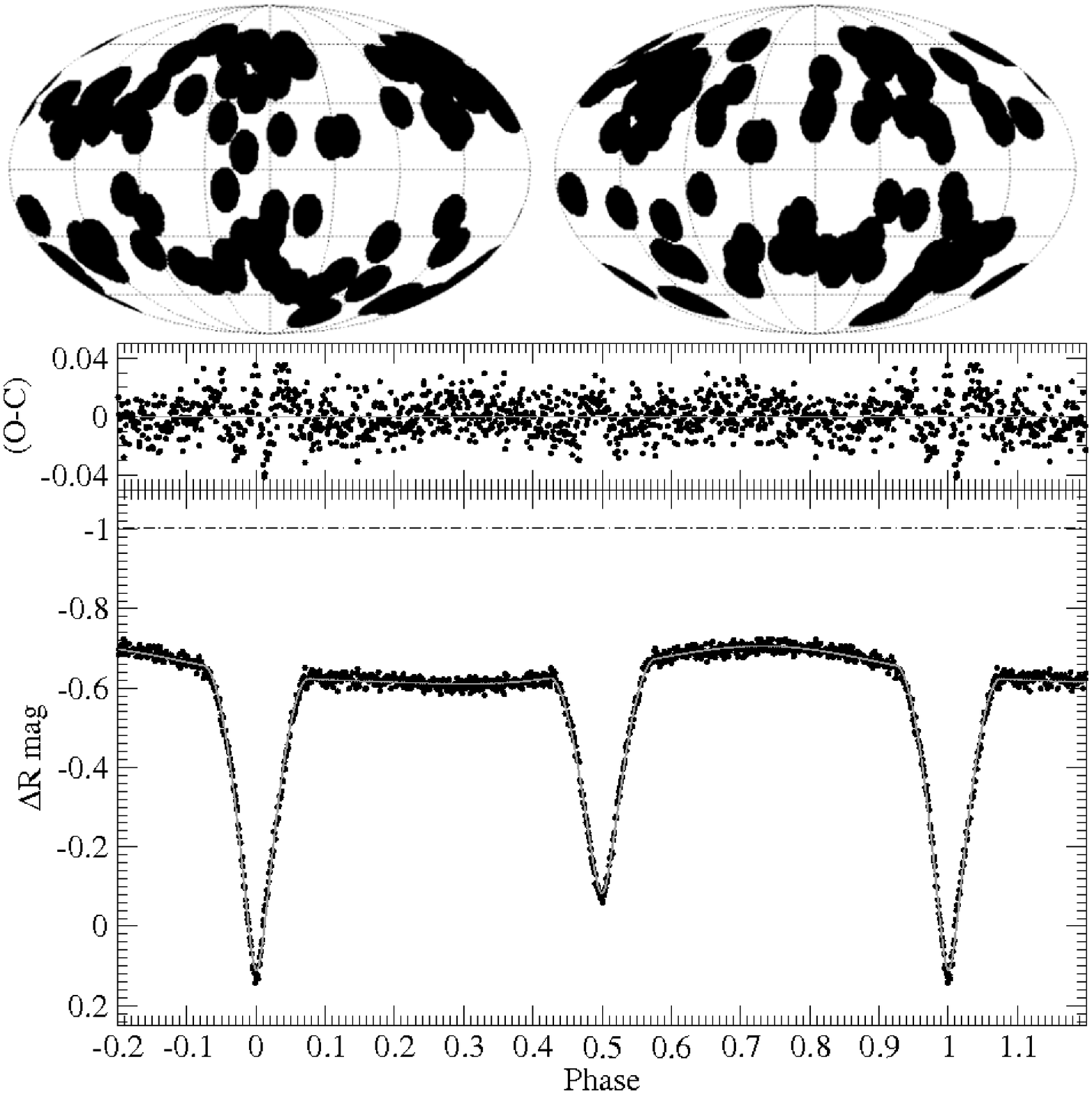}
\includegraphics[width=\columnwidth]{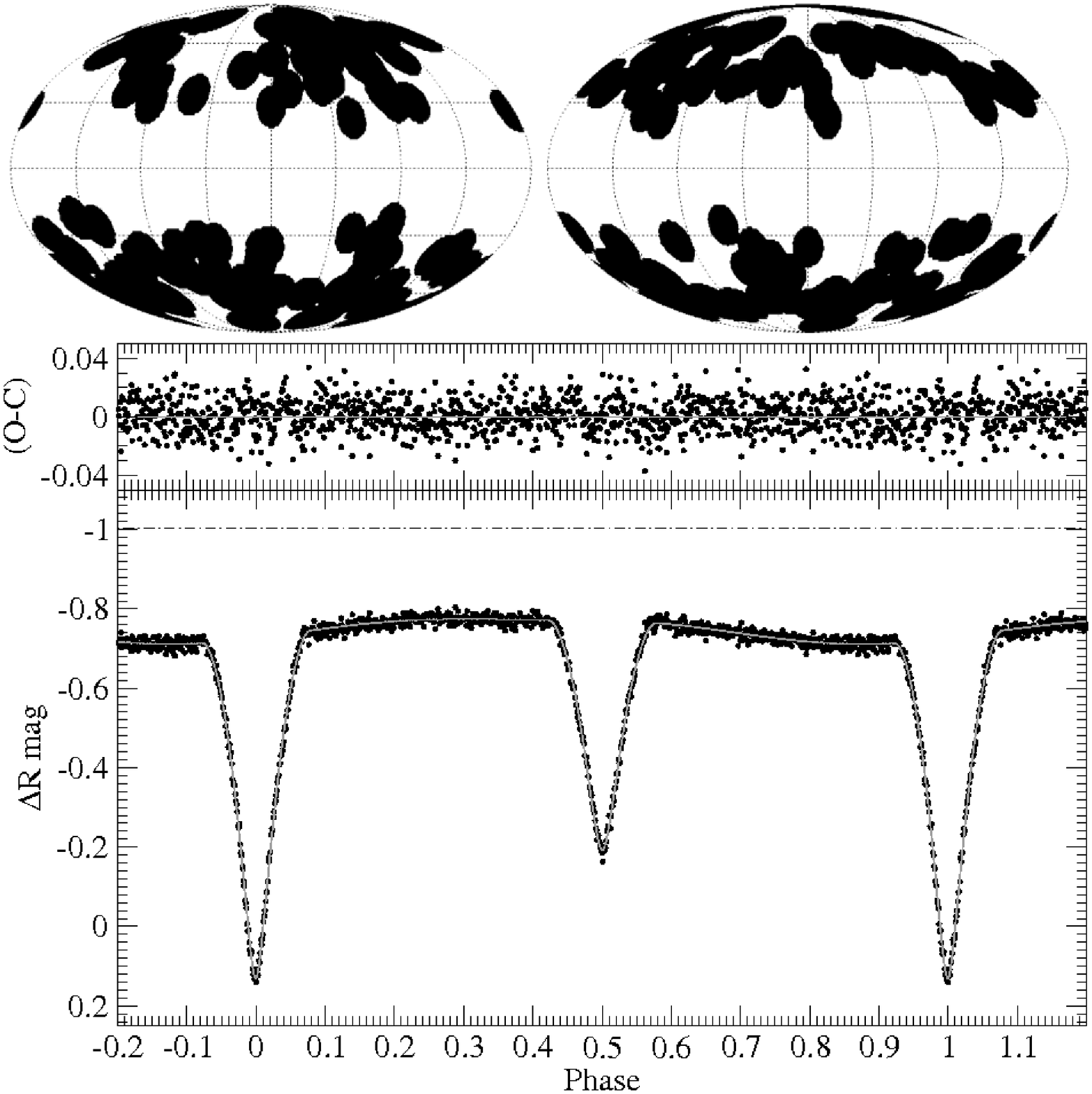}
\caption{Two representations (top panel for Distribution 1 and bottom panel
for Distribution 2) of simulated spotted EB systems (primary component
on the left and secondary on the right) with the resulting light curves. Star
surfaces are represented in a Mollweide projection and the center of each one
corresponds to 0~deg of longitude and 90~deg of co-latitude (equator). Best
fits to the simulated light curves together with the O-C residuals are shown.
Both cases were simulated with $\beta=0.2$. The reference level of the
unspotted light is shown for comparison (dot-dashed line).}
\label{fig:lc_systematics}
\end{figure*}

The simulations described above, were also used to investigate systematic
effects of spots on fits to light curves. A thorough statistical analysis of
the systematic deviations induced by the presence of spots on these parameters
would require fitting all the simulated light curves by treating them as real
observations. Typical light curve analyses involve testing fits for different
parameters, especially when spots are present, since different spot
configurations should be tested. Simultaneous fits of light curve in several
bands are also preferable in order to avoid correlations between parameters
such as limb darkening and radius. Besides, spot parameters are degenerate and
thus the inversion of a light curve to recover the input parameters is an
ill-posed problem. This implies carrying out a slow procedure consisting in
adding spots and evaluating the relevance of the modulation to reach the best
compromise between stability and the quality of the fit to the observations.
For these reasons, fitting all the simulated curves is prohibitively time
consuming. We performed fits to three simulated light curves for each
distribution and for each of the $\beta$ values 0.1, 0.2, 0.3 and 0.4
(33 light curves in total) with the aim of inferring a general trend.
Besides, we focused our attention on the $R$-band to reduce the
computing time. This band is commonly used in the photometric follow-up
of low-mass eclipsing binaries and it is thus representative of light curve
analyses. A possible bias
resulting from the correlation between radius and limb darkening coefficients
should be negligible because the coefficients are computed from the same
theoretical tables \citep{Claret2000} both for simulations and fits. For each
of the $\beta$ values mentioned, we chose as representative spotted light
curves three realizations with modulation amplitudes representative of the
average. 

We added random Gaussian noise of 1\% of the flux to the simulations so that
the typical scatter of light curves is reproduced. We subsequently performed
fits with the WD code. Light curve relevant parameters such as the inclination,
temperature ratio, light ratio and surface potentials were fitted along with
spot properties. Since some of the spot parameters are highly correlated we
started assuming spots at 45~deg latitude and 10\% cooler or hotter than the
photosphere. The free parameters were the longitude ($\phi_{s}$) and size
($r_{s}$) of spots and, in a second step, the latitude ($\theta_{s}$) and
temperature contrast ($\Teff_{,s}/\Teff$). As the usual practice, we included
in the model as many spots as needed to obtain a realistic fit and we tested
the location of spots on primary component, secondary or both at the same time.
All cases yielded satisfactory fits with 1--3 spots on different components. We
found best fits with $\Teff_{,s}/\Teff$ ratios of $\sim 0.9$, thus indicating
that spot groups could be modeled with a smoother temperature contrast than
real. A phase shift was also set as a free parameter to account for
phase corrections due to spots.

\begin{figure*}
\centering
\includegraphics[width=0.7\textwidth]{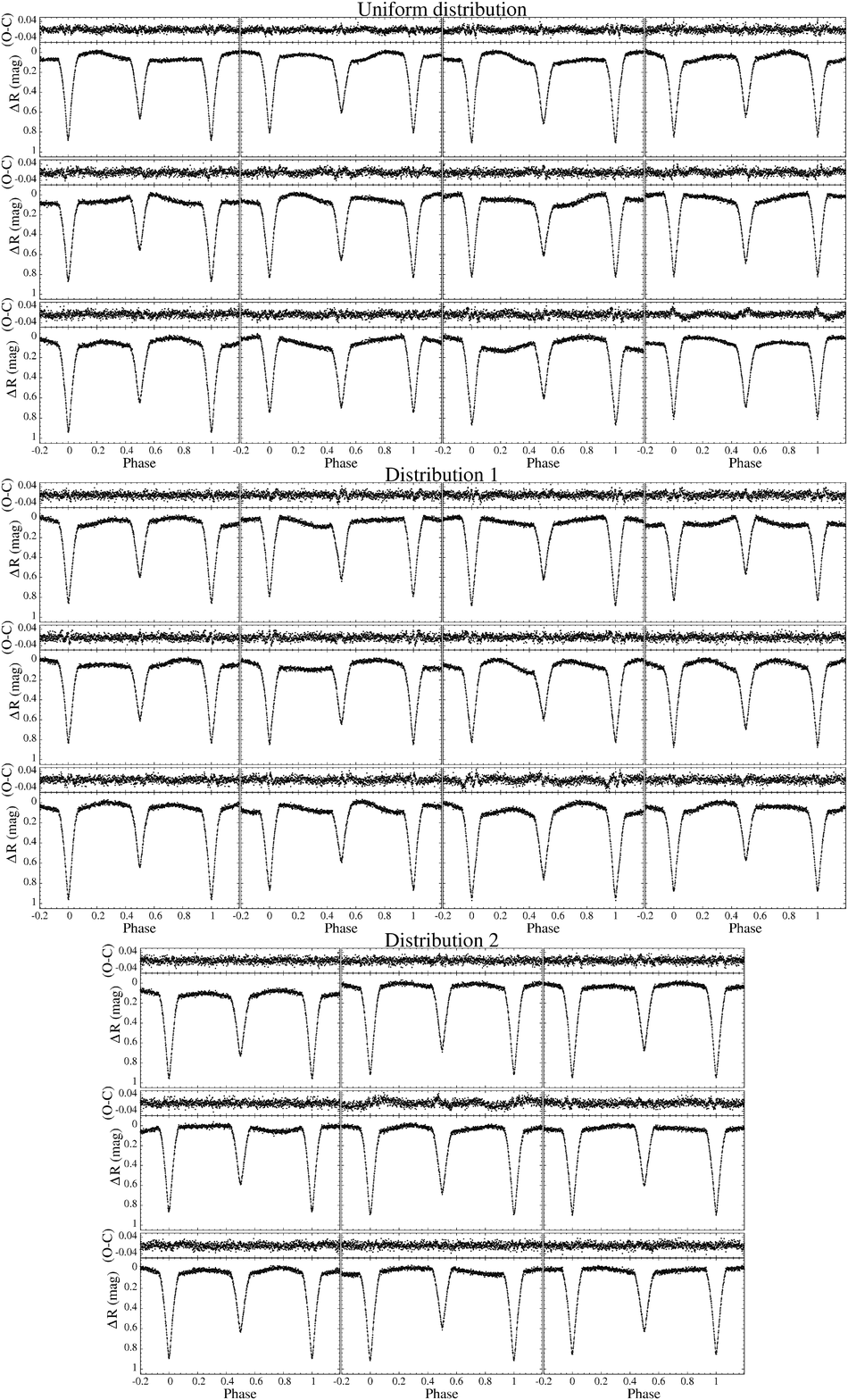}
\caption{Fits to simulated light curves following the procedure explained in
the text. Light curves are vertically shifted to be plotted with the same scale.
Each block corresponds to a different spot distribution as indicated.
Simulations with equal $\beta$ value are arranged in columns, starting from
$\beta=0.1$ to 0.4 in steps of 0.1, except for Distribution 2, for which the
last column corresponds to $\beta=0.3$.}
\label{fig:simulated_LCfits}
\end{figure*}

\begin{figure}[t]
\centering
\includegraphics[width=\columnwidth]{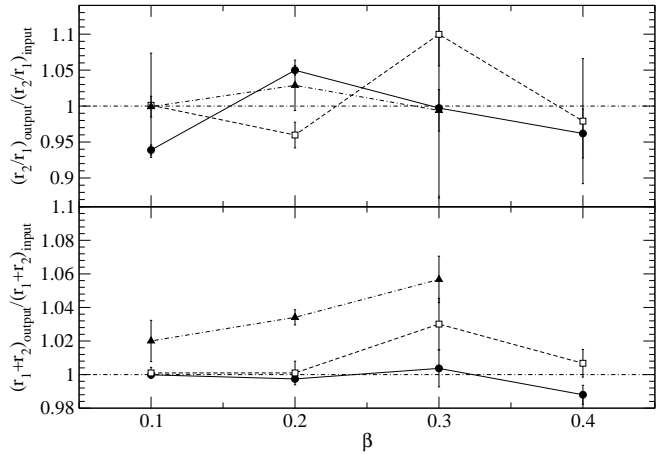}
\caption{Differentials between input parameters to the simulations and those
recovered from fits. The ratio and the sum of the radii are shown on the top
and bottom panels, respectively, for different distributions: uniform (filled
circles), Distribution 1 (open squares) and Distribution 2 (filled triangles).
These values are computed as the average of the results of fits on three
simulated light curves for each distribution. Standard deviations are shown
as error bars.}
\label{fig:diff_R}
\end{figure}

Figure~\ref{fig:lc_systematics} depicts two examples of the fits together with
the configuration of the simulated spots on the stars, and
Figure~\ref{fig:simulated_LCfits} shows all the 33 fits performed. In
Figure~\ref{fig:diff_R} we illustrate the effect of spots on the radius, both
by plotting the mean values and standard deviations (error bars) of the
systematic differences with respect to the input values of the simulations.
The sum and the ratio of radii relative to the semimajor axis were
used to check for systematic effects since these are the parameters that
directly depend on the shape of the eclipses. A clear trend is seen in the case
of Distribution 2, where the sum of the radii of the components (i.e., total
size of the stars) is systematically overestimated by the fits by 2--6\%. For
distributions less concentrated to the pole the deviations induced by spots
seem to be random. In the case of the ratio of radii ratio, the differences
found do not seem to be significant in any of the cases, and especially when 
considering the dispersion. The same lack of systematic trends is found for
other fit parameters such as the inclination, effective temperature ratio, or
light ratio.

The parameters related to radial velocity curves, such as semimajor axis, mass
ratio and systemic radial velocity, were also inspected for deviations by
simulating the corresponding velocity curves.
Fits using the WD code were performed by ignoring the presence of spots and
yielded insignificant differences in the mass ratio and semi-major axis,
in all cases below 1\% and 0.5\%, respectively.
As a consequence, the absolute radius derived for each component of the 
eclipsing binary system is not biased because of the effect of spots on the
semi-major axis, which ultimately defines the scale of the system.

The systematically larger radii found from the distributions with polar spots
have a relatively straightforward geometric interpretation. A projected star
that has a polar cap loses its circular symmetry and its isophotes become 
elongated. This causes the eclipses to widen and the WD code, which assumes a 
Roche geometry (i.e., nearly spherical stars for these well-separated systems)
finds a best fit with a larger stellar radius. In the case of spots more
uniformly distributed in latitude, eclipses are not significantly affected,
thus, systematic deviations appear more random. We tested this by assuming
systems with pure polar caps (i.e., circular and symmetric dark spots covering
both caps) and carrying out fits without considering the presence of spots 
(since no modulations are seen). The results confirm the existence of the
systematic effect on the stellar radii found in the simulations with
Distribution 2. Also, because of the much faster fits, we could test several
input parameters (corresponding to the systems IM~Vir, GU~Boo, YY~Gem, CU~Cnc
and CM~Dra) that yielded comparable results, thus testing the systematic effect
on radius also for different values of relevant binary parameters such as
inclination, relative radius and temperature ratio.
Using these simulations we also tested the influence of the photometric
band by producing light curves in the $BVRIJK$ filters. We found that the
systematic effect on the radius is expectedly smaller in the near-IR bands
because the spot contrast is lower. The radius differences range from $\sim
3.5$\% in $B$ to $\sim 1.2$\% in $K$. The simultaneous fit of several bands
still yields systematic radius differences, corresponding to some average of
those resulting from fits to the individual filter light curves.

The outcome of this battery of tests is that the presence of starspots on the
stellar surface could bias the determination of stellar radii. This is the case
when spots are concentrated towards the poles. Doppler tomography imaging has
revealed a preeminence of polar spots on rapidly rotating low-mass stars, such
as AG~Dor, AB~Dor, LO~Peg \citep[see][for a review]{Strassmeier2009}. YY~Gem is
the only EB system from our sample with published Doppler images
\citep{Hatzes1995}, although no indication of a polar cap was found. However,
polar spots are difficult to detect with Doppler tomography for near edge-on
inclinations, such as the case of YY~Gem. Spectropolarimetric analyses of
low-mass stars showing intense poloidal magnetic fields \citep[see][for a
review]{Donati2009} may provide further evidence of polar spots. If polar caps
are present on low-mass EBs, these could be responsible for up to $\sim 6$\%
(if $\beta=0.3$) of the radius discrepancy between models and observations.
Thus, stellar models should be compared with observed values corrected from
the systematic effect due to polar spots.

\section{Modelling active stars}
\label{sec:models}

\begin{figure*}[t]
\centering
\includegraphics[width=\textwidth]{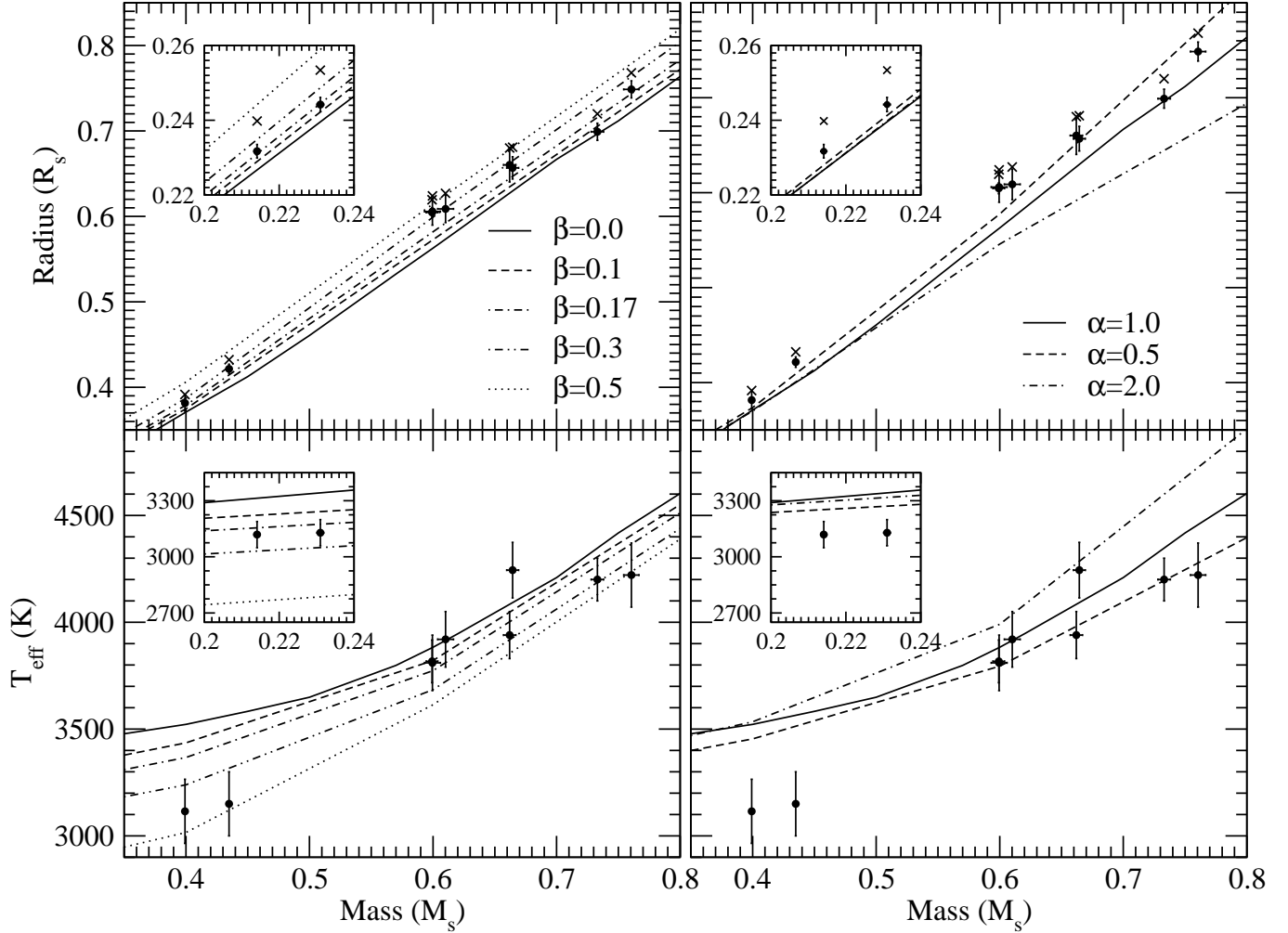}
\caption{Comparison between models and observations for different $\beta$
values and $\alpha=1$ (left), and for different $\alpha$ values and
$\beta=0$ (right). Top panel: $M-R$ relationship. Crosses
indicate the measured radius of the EB components reported in
Table~\ref{tab:best_EBs} and circles represent the values corrected for the
3\% systematic factor described in text and further normalized to an age of
1~Gyr. Bottom: $M-\Teff$ relationship. Circles represent temperatures values
normalized to an age of 1~Gyr. The insets display the case of CM~Dra.}
\label{fig:M_R_corr}
\end{figure*}

As already mentioned in Section~\ref{sec:introduction}, \cite{Chabrier2007}
introduced activity effects in low-mass stellar models by considering two
parameters: the spot blocking factor ($\beta$) and the modification of the
mixing length parameter ($\alpha$). The study of \citet{Chabrier2007} shows
that the effects of the $\beta$ and $\alpha$ parameters are degenerate, i.e.,
the properties of any given system can be reproduced by modifying any of the
two or both, although for completely convective stars, the effect of reducing
the mixing length parameter is very small.

CM Dra offers the possibility of discerning between the effects of activity on
the convective efficiency (parameterized by $\alpha$) and on surface spots
(parameterized by $\beta$). This is because the components of this system
should be fully convective and their structure almost independent of the
mixing length parameter $\alpha$. Thus, we used this system to find the
$\beta$ value that yields the best fit to the properties of the components of
CM~Dra. We compared the theoretical $M-R$ relationships, interpolating between
models with different $\beta$. To do so, we assumed the presence of polar
spots on the components (see Sect. \ref{subsec:lc_systematics}). Different
negative corrections to the radii of the components were used depending on the
value of $\beta$ and in consistency with our simulations in Fig.
\ref{fig:diff_R}. An iterative process was employed until reaching agreement
between the resulting $\beta$ value and the radius correction from light curve
systematics. After iterating, we find that a model with $\beta=0.17 \pm 0.03$
reproduces both components when their radii are downwards corrected by $\sim
3$\%. If spots are $\sim 15$\% cooler than the photosphere, the resulting
$\beta$ translates into ($36 \pm 6$)\% of the star surface covered by spots, a
value that is in agreement with findings from Doppler imaging for other systems
such as HK~Aqr \citep{Barnes2001}. Note, however, that spot coverages from
Doppler tomography represent a lower limit because of the limited sensitivity
of the technique to small-sized spots or low contrast temperatures.

Abundant X-ray observations have revealed a clear relation between magnetic
activity and rotation, indicating that rapidly rotating stars are more active
than slower rotators up to a certain limit where saturation is reached.
\cite{Pizzolato2003} showed that M-dwarfs with rotation periods below
$\sim 10$~days show saturation of their X-ray emission, therefore we could
assume that binaries in our sample are all in this regime of saturated coronal
activity. The physical mechanism of saturation is not well understood yet, but
one of the feasible explanations could be that saturation is reached when the
entire stellar corona is full of active regions. This may imply that the
$\beta$ parameter is similar for stars that are in the saturated regime. 

At this point, we make two assumptions to proceed with the analysis: {\em (1)}
that spots on active EBs are mostly polar and {\em (2)} that saturated systems
have similar spot coverages (i.e., that $\beta$ is roughly independent of mass
for very active systems). With these assumptions and following the same
procedure applied above for CM Dra, we find that the overall sample of EB
systems is reproduced within 1$\sigma$ of their error bars with $\beta=0.20 \pm
0.04$, i.e., ($42 \pm 8$)\% of spot coverage.

The left panel of Figure \ref{fig:M_R_corr} shows a comparison of mass and
radius observations with models assuming different $\beta$ values. The radii of
the EB systems shown in the figure were corrected considering two effects.
First, a 3\% systematic difference resulting from the light curve analysis.
Second, an age effect on the stellar radii to put all systems at a normalized
age of 1~Gyr (ages in Table \ref{tab:best_EBs}, except for NGC2204-S892 where
no age is available and no correction was made) using the models of
\cite{Baraffe1998}. The evolutionary radius offsets are always well below the
1\% level. This allowed us to plot Figure \ref{fig:M_R_corr} and compare all
EBs with a theoretical model with the same age.

To serve as a reference, if we relax the polar spot assumption, i.e., we do not
correct for the systematic effect of polar spots on light curves, the best
model reproducing CM~Dra gives $\beta=0.37 \pm 0.04$, which means more than
75\% of the surface of each component covered by spots. This seems to be a
rather high value of spot coverage when compared with the results of Doppler
imaging. If we do not consider the saturation hypothesis and assume that each
system could be reproduced by different $\beta$ factors, we find in systems
such as YY~Gem or GU~Boo over half of the surface must be spotted. Note that
the spot coverage estimates depend on the spot contrast, which is set to $\sim
500$~K in this work.

\begin{figure}[t]
\centering
\includegraphics[width=\columnwidth]{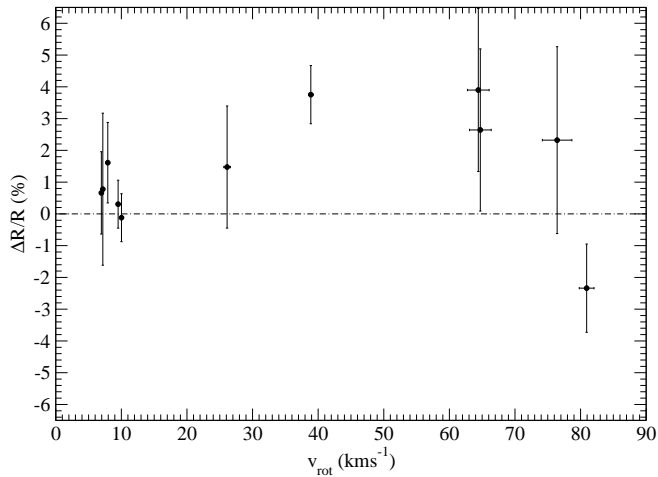}
\caption{Residual discrepancy between observations and the model with
$\beta$=0.17 that best fits the case of CM~Dra, which is not affected by
variations of the mixing length parameter $\alpha$.}
\label{fig:dr_rot}
\end{figure}

Some remaining radius differences are still apparent in Figure
\ref{fig:M_R_corr} when considering a $\beta$ value of 0.17. Systems such as
YY~Gem or GU~Boo have observed radii well over 1$\sigma$ of the model
predictions. It is interesting to note that these are the systems with the
fastest rotational velocity and perhaps be an indication of additional effects
of rotation and/or magnetic activity not accounted for in the $\beta$ analysis.
The right panel of Figure \ref{fig:M_R_corr} shows that for the more massive
M-type EBs ($<0.6$~M$_{\odot}$) the additional differences could be explained
by reducing the mixing length parameter $\alpha$. Figure~\ref{fig:dr_rot}
depicts the differences between the $\beta=0.17$ model and observations (after
correction for polar spots and normalized to 1~Gyr) as a function of the
rotational velocity of each star (computed assuming synchronization). Although
not significant, a general trend of increasing differences for increasing
rotational velocities seems to be present. The exception is NGC2204-S892 (two
rightmost points on the plot), which is the shortest period binary. But note
that no determination of the age and metallicity of this EB system is available
and this could be falsifying the analysis. The tentative trend may indicate
that, while the spot effect (via the $\beta$ parameter) is a clear contributor
to explaining the differences between observation and theory, the loss of
efficiency of convective energy transport (related with magnetic fields and
thus possibly with rotation rate) may also be at play, explaining up to 4\% of
the radius difference.

As stated in Section~\ref{sec:sample}, absolute effective temperatures of EB
components were not used to constrain model parameters because they are not
determined independently of calibrations as masses and radii. Nevertheless we
checked the consistency of the spot scenario with temperatures. The bottom
panels of Figure~\ref{fig:M_R_corr} show the comparison between models and
observations in the $M-\Teff$ plane for different $\alpha$ and $\beta$
values. The temperatures represented on these plots have been also normalized
to an age of 1~Gyr using the models of \cite{Baraffe1998} although differences
amount for less than 10~K in all cases. The model with
$\beta=0.17$ also reproduces the $M-\Teff$ relationship of EBs within the
errors, with the exceptions of IM~Vir and CU~Cnc, whose temperatures may be
affected by a circumstellar dust disk. These figures also show that both
models with lower $\alpha$ and the model with $\beta=0.17$ reproduces all
systems but CM~Dra, due to the negligible effect of mixing length on
completely convective stars. Thus, the conclusions from the $M-R$ analysis are
also consistent with the effective temperature comparison.

\section{Conclusions}
\label{sec:conclusions}

After evaluating different scenarios to explain the discrepancies
reported between observation and theory for low mass stars, a rather complex
picture is now emerging since no single effect can account for the full
size of the differences. Our analysis shows that the $\sim 5-10$\% radius
discrepancy in M-type EBs is explained by combining three factors: {\em (1)}
A systematic overestimation of the stellar radius in EB light curve analysis
caused by the presence of polar spots with the subsequent loss of circular
symmetry (amounting to about $\sim 3$\%); {\em (2)} an increase of stellar
radii to compensate for the loss of radiative efficiency because of starspots
(reproduced by models considering a $\beta$ parameter of 0.17 that explains 2\%
of the radius difference); and {\em (3)} an increase of the radius caused by
the lower convective efficiency in fast rotators with supposedly strong
magnetic fields (reproduced by models through changing the $\alpha$ parameter)
and accounting for 0--4\% of the radius difference. This scenario also
reproduces the $M-\Teff$ distribution of EBs although effective temperatures
are not as fundamentally determined as radii.

Our overall analysis ensures self consistency by comparing with the constraints
set by the observations. Predicted spot coverages and simulated light curves
are found to reproduce the modulation amplitudes in the real light curves of
the EB systems. Also the $\sim 35$\% surface coverage is found to be compatible
with results from Doppler tomography of active stars. Even relatively inactive
systems such as some hosts of transiting exoplanets are found to have
significant spot coverages when applying transit mapping procedures
\citep{Pont2007,Silva-Valio2008}.

It is important to note that our results are based on two main hypotheses: {\em
(1)} That {\em all} of these active EB components have spots that
preferentially occupy locations close to the pole (with consequent effect on
the light curve analysis) in line with results from Doppler imaging; and {\em
(2)} that {\em all} EB components have similar surface spottedness caused by
their saturated activity regime. The first hypothesis relaxes the classical
discrepancy with models to a lower 2--7\%, and the second hypothesis ensures
that $\beta$ will roughly be ``universal'' for these strongly active systems.
In this framework, the system CM Dra is key in reaching the conclusions above.
This is because it allows us to separate the effects of the two free
parameters, namely $\beta$ and $\alpha$, since the components of this system
are expected to be fully convective and little affected by changes in the
mixing length parameter.

The analysis of low-mass stellar evolution models clearly shows that activity
plays an important role on the structure of stars, and that when considering
activity effects on stellar structure, models can reproduce the observed values
of radii and effective temperatures of low-mass stars. Accurate masses and
radii of new low-mass systems coming from missions such as CoRoT and Kepler
should improve the statistics and better define the scenario proposed above,
especially if comparable active and non-active EB systems are found. Doppler
imaging of the EBs analyzed or new methods of mapping stars using
interferometry \citep{Monnier2007} should also give indications on the spot
distribution on the stellar surfaces. All these efforts are valuable to
understand the vast majority of the stars that populate the Galaxy and,
notably, have an impact in the area of exoplanet research, in which planet
properties are directly linked to our knowledge of the host star.

\acknowledgements
We are grateful to J. F. Donati for useful discussions and to the 
referee for helpful suggestions. JCM, IR and CJ acknowledge support from the
Spanish Ministerio de Ciencia e Innovaci\'on via grants AYA2006-15623-C02-01,
AYA2006-15623-C02-02, AYA2009-06934, and AYA2009-14648-C02-01. This work was
partly supported by the French ANR ``Magnetic Protostars and Planets'' (MAPP)
project and by the ``Programme National de Physique Stellaire'' (PNPS).


\begin{thebibliography}{}
\bibitem[Andersen(1991)]{Andersen1991} Andersen, J.\ 1991, \aapr, 3, 91
\bibitem[Bahcall et~al.(2005)]{Bahcall2005} Bahcall, J.\ N., Basu, S., Pinsonneault, M., \& Serenelli, A.\ M.\ 2005, \apj, 618, 1049
\bibitem[Bahcall et~al.(2004)]{Bahcall2004} Bahcall, J.\ N., Serenelli, A.\ M., \& Pinsonneault, M.\ 2004, \apj, 614, 464
\bibitem[Baraffe et~al.(1998)]{Baraffe1998} Baraffe, I., Chabrier, G., Allard, F., \& Hauschildt, P.\ H.\ 1998, \aap, 337, 403
\bibitem[Barnes \& Collier Cameron(2001)]{Barnes2001} Barnes, J.\ R., \& Collier Cameron, A.\ 2001, \mnras, 326, 950
\bibitem[Berger et al.(2006)]{Berger2006} Berger, D.~H. et al.\ 2006, \apj, 644, 475
\bibitem[Casagrande et al.(2008)]{Casagrande2008} Casagrande, L., Flynn, C., \& Bessell, M.\ 2008, \mnras, 389, 585
\bibitem[Chabrier \& Baraffe(1997)]{Chabrier1997} Chabrier, G., \& Baraffe, I.\ 1997, \aap, 327, 1039
\bibitem[Chabrier et al.(2007)]{Chabrier2007} Chabrier, G., Gallardo, J., \& Baraffe, I.\ 2007, \aap, 472L, 17
\bibitem[Claret(2000)]{Claret2000} Claret, A.\ 2000a, \aap, 363, 1081
\bibitem[Demory et al.(2009)]{Demory2009} Demory, B.-O., et al.\ 2009, \aap, 505, 205
\bibitem[Donati \& Landstreet(2009)]{Donati2009} Donati, J.-F., \& Landstreet, J.\ D.\ 2009, \araa, 47, 333
\bibitem[Granzer et al.(2000)]{Granzer2000} Granzer, T., Sch\"ussler, M., Caligari, P., \& Strassmeier, K.~G.\ 2000, \aap, 355, 1087
\bibitem[Hatzes(1995)]{Hatzes1995} Hatzes, A.\ P.\ 1995, IAU Symposium, 176, 90
\bibitem[Jeffers et al.(2007)]{Jeffers2007} Jeffers, S.\ V., Donati, J.\ F., \& Collier Cameron, A.\ 2007, \mnras, 375, 567 
\bibitem[L\'opez-Morales(2007)]{LopezMorales2007} L\'opez-Morales, M.\ 2007, \apj, 660, 732
\bibitem[L\'opez-Morales \& Ribas(2005)]{LopezMorales2005} L\'opez-Morales, M., \& Ribas, I.\ 2005, \apj, 631, 1120
\bibitem[Mazeh(2008)]{Mazeh2008} Mazeh, T.\ 2008, EAS Publications Series, 29, 1 (astro-ph/0801.0134)
\bibitem[Monnier et al.(2007)]{Monnier2007} Monnier, J.\ D., et al.\ 2007, Science, 317, 342
\bibitem[Morales et al.(2008)]{Morales2008} Morales, J.\ C., Ribas, I., \& Jordi, C.\ 2008, \aap, 478, 507
\bibitem[Morales et al.(2009a)]{Morales2009a} Morales, J.\ C., et al.\ 2009a, \apj, 691, 1400
\bibitem[Morales et al.(2009b)]{Morales2009b} Morales, J.\ C., Torres, G., Marschall, L.\ A., \& Brehm, W.\ 2009b, \apj, 707, 671
\bibitem[Pizzolato et al.(2003)]{Pizzolato2003} Pizzolato, N., Maggio, A., Micela, G., Sciortino, S., \& Ventura, P.\ 2003, \aap, 397, 147
\bibitem[Pont et al.(2007)]{Pont2007} Pont, F., et al.\ 2007, \aap, 476, 1347
\bibitem[Ribas(2003)]{Ribas2003} Ribas, I. 2003, \aap, 398, 239
\bibitem[Ribas(2006)]{Ribas2006} Ribas, I.\ 2006, in ASP Conf. Ser. 55, Astrophysics of Variable Stars, ed. C.~Sterken \& C.~Aerts (San Francisco: ASP), 349
\bibitem[Ribas et al.(2008)]{Ribas2008} Ribas, I., Morales, J.~C., Jordi, C., Baraffe, I., Chabrier, G., \& Gallardo, J.\ 2008, \memsai, 79, 562
\bibitem[Rozyczka et al.(2009)]{Rozyczka2009} Rozyczka, M., Kaluzny, J., Pietrukowicz, P., Pych, W., Mazur, B., Catelan, M., \& Thompson, I.\ B.\ 2009 (astro-ph/0910.2543)
\bibitem[Strassmeier(2009)]{Strassmeier2009} Strassmeier, K.\ G.\ 2009, \aapr, 17, 251
\bibitem[Silva-Valio(2008)]{Silva-Valio2008} Silva-Valio, A.\ 2008, \apjl, 683, 179
\bibitem[Torres et al.(2009)]{Torres2009} Torres, G, Andersen, J., \& Gim\'enez, A.\ 2009, \aapr, 18, 67
\bibitem[Torres et al.(2006)]{Torres2006} Torres, G., Lacy, C.\ H., Marschall, L.\ A., Sheets, H.\ A., \& Mader, J.\ A.\ 2006, \apj, 640, 1018
\bibitem[Torres \& Ribas(2002)]{Torres2002} Torres, G., \& Ribas, I. 2002, \apj, 567, 1140
\bibitem[Washuettl \& Strassmeier(2001)]{Washuettl2001} Washuettl, A. \& Strassmeier, K.\ G.\ 2001, \aap, 370, 218
\bibitem[Wilson \& Devinney(1971)]{WD1971} Wilson, R.~E., \& Devinney, E.~J.\ 1971, \apj, 166, 605
\end{thebibliography}
\end{document}